\documentclass[10pt,pra,aps,twocolumn,notitlepage,nofootinbib,superscriptaddress]{revtex4-1}

\usepackage [english]{babel}
\usepackage [utf8]{inputenc}
\usepackage [T1]{fontenc}
\usepackage{graphicx}
\usepackage{amsmath}
\usepackage{amssymb}
\usepackage{bbold}
\usepackage[dvipsnames]{xcolor}
\usepackage{natbib}
\usepackage{subfigure}
\usepackage[colorinlistoftodos, shadow,textsize= footnotesize]{todonotes}
\usepackage[linktocpage=true,colorlinks=true,citecolor=magenta]{hyperref}

\usepackage{empheq}
\usepackage{tikz}
\usetikzlibrary{decorations.pathmorphing, positioning, shapes.geometric}

\usepackage[xcolor]{changes}
\definechangesauthor[color= orange]{GS}
\definechangesauthor[color=blue]{VS}
\definechangesauthor[color= magenta]{FSR}

\newcommand{\tr}{\mathrm{tr}}

\newcommand{\rref}[1]{(\ref{#1})}

\newcommand{\ket}[1]{|#1\rangle}

\begin{document}
\date{\today}

\title{Diffraction-induced entanglement loss of orbital-angular-momentum states}
\author{Giacomo Sorelli} 
\affiliation{Physikalisches Institut, Albert-Ludwigs-Universit\"at Freiburg, Hermann-Herder-Stra\ss e 3, D-79104 Freiburg, Germany}
\author{Vyacheslav N. Shatokhin}
\affiliation{Physikalisches Institut, Albert-Ludwigs-Universit\"at Freiburg, Hermann-Herder-Stra\ss e 3, D-79104 Freiburg, Germany}
\author{Filippus S. Roux}
\affiliation{National Metrology Institute of South Africa, Meiring Naud\'e Road, Brummeria, Pretoria, South Africa}
\affiliation{School of Physics, University of the Witwatersrand, Johannesburg 2000, South Africa}
 \author{Andreas Buchleitner}
\affiliation{Physikalisches Institut, Albert-Ludwigs-Universit\"at Freiburg, Hermann-Herder-Stra\ss e 3, D-79104 Freiburg, Germany}
\begin{abstract}
We provide an analytical expression for the entanglement decay of initially maximally entangled orbital angular momentum 
bi-photon states, 
when scattered off an obstruction. We show that the 
decay is controlled by the 
diffraction-induced mutual overlap between the diffracted
field modes, and quantify its dependence on 
the size and position of the obstruction.
\end{abstract}
\maketitle
\section{Introduction}
Photons with a helical phase front have a definite orbital angular momentum (OAM) 
$l\hbar$, with $l$ an arbitrary integer \cite{AndrewsBook}. 
The arguably most 
attractive feature of this spatial degree of freedom is that it spans an  unbounded, discrete Hilbert space and, thus, can be used 
to encode high-dimensional, possibly entangled (qudit) states \cite{KrennPNAS2014} which are considered as a resource to achieve increased channel capacities \cite{WangNatPhot2012}, and to enhance the security of quantum communication protocols \cite{MafuPRA2013,MirhosseiniNJP2016}.

However, the information encoded in OAM photonic states is very fragile with respect to disturbances 
along the transmission path. For instance, in free space, 
the scattering of photons on random inhomogeneities of the refractive-index of air, as generic in turbulent atmosphere,
results in the crosstalk (coupling) amongst distinct spatial OAM modes \cite{Anguita:08,TylerOptLett09} and leads 
to the (unavoidable and irreversible) decay of 
OAM entanglement \cite{raymer06,RouxPRA2011,IbrahimPRA2013,LeonhardPRA2015,RouxPRA2016}. 
Another, distinct mechanism of entanglement degradation is due to
diffractive effects as induced by physical obstructions. It is in these cases suggestive 
that a suitable choice of the field modes used to encode the OAM state which is to be transmitted may allow to compensate for diffractive effects.
Indeed, recent experiments on the entanglement evolution of bipartite OAM entangled states which are diffracted upon a circular obstruction provide evidence that measurements in the Bessel-Gaussian (BG) rather than in the Laguerre-Gaussian (LG) basis allow for the reduction of diffraction-induced losses of entanglement. This observation was qualitatively attributed to the known self-healing property of BG modes \cite{McLarenNatPhys2014}.

In our present contribution, we offer a general and quantitative theoretical treatment of diffraction-induced entanglement decay 
in terms of the concomitant overlap between the diffracted modes, and will see that this mutual overlap is in general significantly smaller for BG as compared to LG modes.
After a brief recollection of basic properties of LG and BG modes in Sec.~\ref{modes}, Sec. \ref{prop} presents our treatment of the 
diffraction of a single photon on an obstruction.
Sec. \ref{EnCo} generalizes the diffraction problem to entangled
bi-photons, to derive the above result for twin-photon entanglement past an obstacle, before Sec. \ref{conc} concludes our work.

\section{Laguerre-Gaussian  and Bessel-Gaussian modes}
\label{modes}
Let us consider a scalar monochromatic wave $\tilde{\psi}(x,y,z,t)=\psi(x,y,z) e^{-i\omega t}$ propagating along the positive $z-$direction. Its spatial part obeys the Helmholtz equation, which in the paraxial approximation turns into the homogeneous parabolic equation \cite{goodman},
\begin{equation}
2ik\frac{\partial{\psi}}{\partial{z}} = \nabla^2_T \psi ,
\label{parabolic}
\end{equation}
where $\nabla^2_T=\partial^2/\partial x^2 + \partial^2/\partial y^2$ is the transverse Laplacian  and $k = 2\pi /\lambda$ is the wave number, with $\lambda$ the wave length. 

A cylindrically symmetric eigensystem of  Eq. (\ref{parabolic}) is formed by Laguerre-Gaussian (LG)  modes. The latter
are characterized by two discrete indices: the azimuthal index $l=0,\pm 1,\pm 2,\ldots$ associated with the OAM $\hbar l$ per photon populating the mode, and the radial index $p=0,1,2,\ldots$ which 
fixes the radial intensity distribution in the transverse plane, with 
$p+1$ concentric rings of local intensity maxima.
LG modes were the first OAM-carrying light beams to be studied \cite{AllenPRA1992}, and they are commonly used to investigate OAM entanglement \cite{MairLettNature2001}.

Another cylindrically-symmetric set of solutions to Eq. \eqref{parabolic}
is provided by Bessel-Gaussian (BG) modes \cite{GoriOptComm1987} that are a physical approximation of Bessel beams. The latter are formal, non-diffracting 
solutions of the Helmholtz equation  formed by superpositions of plane waves whose wave vectors lie on a cone \cite{DurninPRL1987,DurninOpt1987}, though require infinite energy for their creation, as their intensity has to be reproduced at any $z>0$. Note, however that BG modes are neither complete nor orthogonal in the entire transverse space, but only in the subspace associated with the OAM degree of freedom. The latter property qualifies them as a suitable basis for OAM-encoded information transmission.
 
BG modes \cite{GoriOptComm1987} are specified by two parameters: The discrete azimuthal index $l$, which, as well as for LG modes, defines the OAM, and the (continuous) radial wave number $k_\rho\geq 0$, which characterizes the mode's radial structure. By changing $k_\rho$, the transverse intensity distribution is continuously transformed from a Gaussian  ($k_\rho = 0$) to a multi-ring intensity distribution, which is characteristic of BG beams. 
Since BG modes are renowned for their self-healing property, i.e., the ability to reconstruct after encountering an obstruction \cite{BouchalOC1998,BouchalOC2002,LitvinOC2009}, they are potentially interesting carriers of OAM entanglement \cite{McLarenOSA2012}.

\section{Diffraction of a single photon on an opaque screen}

We now address the modification of a given photonic input state upon scattering off an obstruction. We first consider the case of a single 
photon, and will use these results to model the fate of a bi-photon state in the subsequent Section.
\label{prop}
\subsection{Input states}

Single photon input states with well-defined OAM $\hbar l_0$ are accommodated by the LG mode
\begin{equation}
u^{0l_0}_{LG}(x,y,z=0) = \mathcal{N}_{LG} \left(\rho/w_{LG}\right)^{|l_0|}e^{i l_0 \phi} e^{-\rho^2/w_{LG}^2},
\label{LG}
\end{equation}
with $p=0$, and by the 
BG mode 
\begin{equation}
u^{\kappa l_0}_{BG}(x,y,z=0) = \mathcal{N}_{BG} J_{l_0}(\kappa \rho) e^{i l_0 \phi}e^{-\rho^2/w_{BG}^2},
\label{BG}
\end{equation}
with $k_\rho=\kappa$, where $\mathcal{N}_{LG}$ and $\mathcal{N}_{BG}$ are normalization constants 
(their explicit form is irrelevant for our subsequent analysis), $\rho=\sqrt{x^2+y^2}$ the radius, $\phi$ the azimuthal angle in the $x-y$ plane, $J_{l_0}(x)$ the Bessel function of order $l_0$, 
and $w_{LG}$, $w_{BG}$ the mode waists. $z=0$ is chosen such as to identify the position of the beam 
waists, where, for simplicity, we also place the obstacle.

Anticipating our results on how an obstruction affects the input modes, the following remark is in order:
As follows from Eqs. (\ref{LG}), (\ref{BG}), the LG (BG) modes have single-(multi-)ring intensity distributions that depend on their beam widths and azimuthal indices. Therefore, an obstruction of a given radius $a$ (which we set equal to the experimental value chose in \cite{McLarenNatPhys2014}), in general will screen out different structural elements of incident beams with distinct azimuthal indices. 
However, for BG modes, because of the presence of the Bessel function in Eq. (\ref{BG}), the transverse structure is such that intensity is spread over multiple rings and the fraction of intensity obscured by the obstacle is almost independent of $l_0$. Consequentely, a constant beam waist $w_{BG}$ given by the experimental value \cite{McLarenNatPhys2014} is used in the following. In contrast, the intensity of LG modes is distributed over a single ring with the $l_0-$dependent radius $\rho = \sqrt{l_0/2} w_{LG}$ (see Eq. \eqref{LG}).
If -- to ensure that the maximum of the input intensity distribution be covered by the obstacle placed on the beam axis -- we impose $\rho=0.8 a$, the beam waist needs to be chosen $l_0-$dependent, according to $w_{LG}=0.8a\sqrt{2/l_0}$. 

By comparison of modes with single (LG) and multi ring (BG) intensity distributions, we aim at a better understanding of how the two different radial structures affect the transmission of OAM states across obstructed paths.

Hereafter, we denote single photon states prepared in the spatial modes \eqref{LG} and \eqref{BG} as $|u_{LG}\rangle:=|0,l_0\rangle$ and $|u_{BG}\rangle:=|\kappa,l_0\rangle$, respectively.

\subsection{Boundary conditions}
\label{sec:boundary}
We seek the solution $\psi(x,y,z)$ of \eqref{parabolic} for $z>0$, 
for a single photon state of the spatial input mode $u(x,y,z)$ (we skip the labels $LG$ or $BG$), diffracted by an obstruction located at $z=0$. Following the scenario of the experiment \cite{McLarenNatPhys2014}, we assume a circular shape of the obstacle, which is also convenient for our theoretical analysis, owing to its symmetry. Note, however, that this does not imply a fundamental restriction: the diffracted mode $\psi(x,y,z)$ 
can be numerically inferred (see Sec. \ref{sec:basic}) for arbitrary geometries.

To account for the impact of the obstacle, we impose the {\it modified} Kirchhoff boundary conditions \cite{goodman,FischerOptExpr2007},
\begin{equation}
\psi(x,y,z=0) = u(x,y,z=0)t(x-d,y),
\label{bc}
\end{equation}
where 
\begin{equation}
t(x-d,y) = 1 -\exp \left\lbrace - \left[\frac{(x-d)^2 + y^2}{a^2}\right]^m \right\rbrace,
\label{t}
\end{equation}
is the obstacle's transmission function, with $d$ and $a $ its
shift\footnote{Due to the rotational symmetry of the problem,
there is no loss of generality in our choice of the displacement $d$
along the $x$-axis.} with respect to the beam axis and 
its radius, respectively, and $m$ a positive integer. In our simulations, we choose  $m=12$, and the thus defined super-Gaussian on the right hand side of Eq. (\ref{t}) serves to smoothen edge effects as encountered \cite{FischerOptExpr2007} for Kirchhoff boundary conditions, $t(x-d,y) = \Theta((x-d)^2+y^2 - a^2)$, with $\Theta(x)$ the Heaviside function. 

Since part of the incident mode is blocked by the obstruction, the diffracted mode still needs to be renormalized:
\begin{equation}
\psi(x,y,z) := \frac{\psi(x,y,z)}{(\int\int dx dy |\psi(x,y,0)|^2)^{1/2}}.
\label{norm_psi}
\end{equation}  
Eq. (\ref{norm_psi}) ensures that $\int\int dx dy |\psi(x,y,z)|^2=1$ for any $z\geq 0$. 
 
\subsection{Basic properties of the diffracted wave} 
\label{sec:basic}
\begin{figure*}
\includegraphics[width=0.9\linewidth]{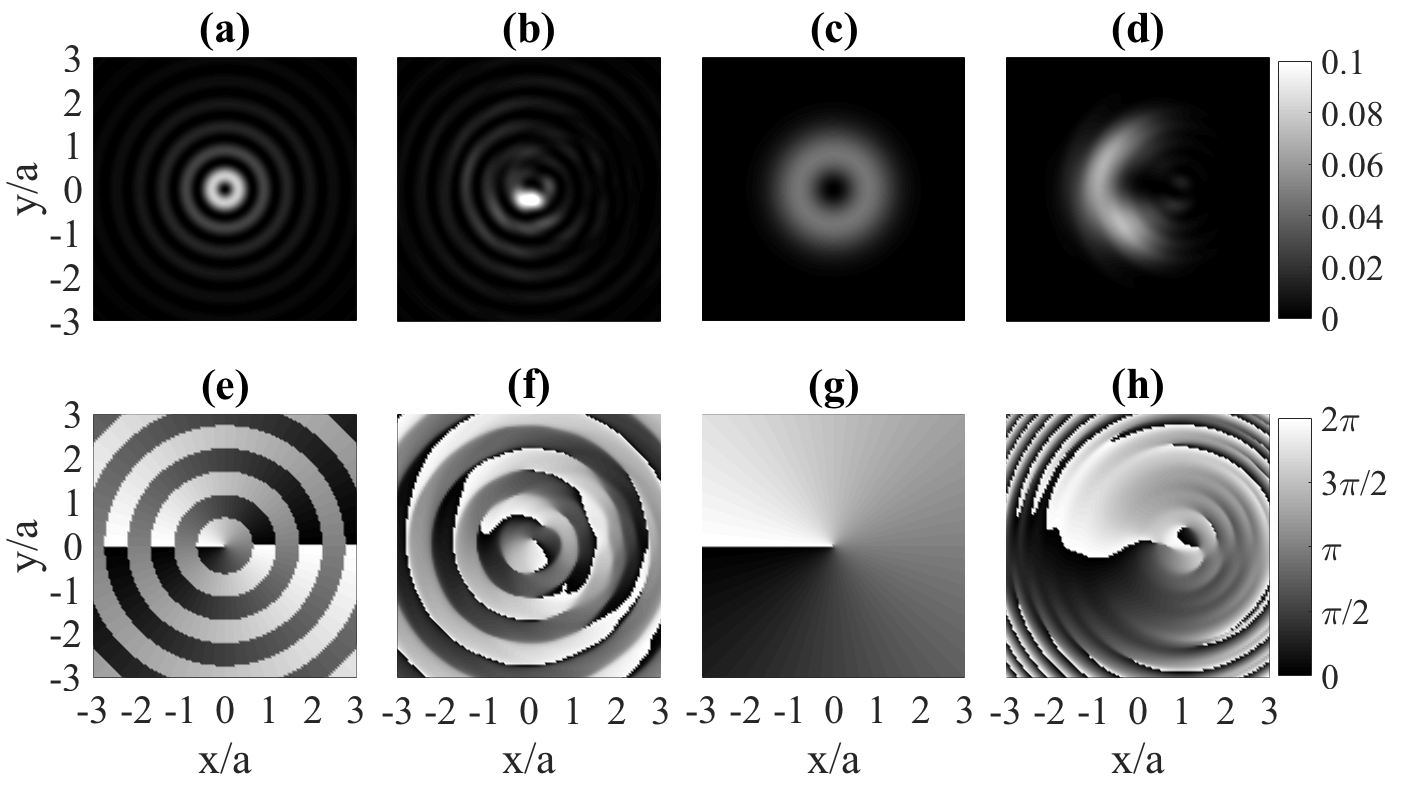}
\caption{Intensity (a-d) and phase (e-h) distribution of BG (a,b,e,f) and LG (c,d,g,h) modes with azimuthal index $l_0=1$. Subplots (a, e) and (c, g) correspond to the unperturbed LG and BG modes $u(x,y,z=0)$, respectively [see Eqs. \eqref{LG} and \eqref{BG}], whereas subplots (b, f, d, h) represent diffracted waves $\psi(x,y,z)$ at $z=50$ mm behind a circular obstacle with the transmission function $t(x-d,y)$ [see Eq. \eqref{t}] parametrized by $a =d = 200\; \mu$m  and $m=12$.
The specific parameter values employed to produce the present patterns are:
$p=0$, $\kappa=30$ mm$^{-1}$, $w_{LG}= 226 \; \mu$m, $w_{BG}= 1$ mm, $\lambda= 710$ nm, and, except for $w_{LG}$, remain fixed throughout this work.}
\label{InPhase}
\end{figure*}

An exact numerical solution of the boundary value problem defined by Eqs.~\eqref{parabolic},~\eqref{bc}-\eqref{norm_psi}, for arbitrary parametrisation of the unperturbed 
beams \eqref{LG}, \eqref{BG} and of the transmission function \eqref{t}, can be obtained with the help of the {\it angular-spectrum propagator}, well-known in Fourier optics  \cite{goodman},
\begin{equation}
\psi (x,y,z) = \mathcal{F}^{-1}\lbrace T(k_x,k_y,z) \mathcal{F}\left[\psi (x,y,z=0)\right]\rbrace,
\label{angspec}
\end{equation} 
where ($\mathcal{F}^{-1}$) $\mathcal{F}$ indicates the (inverse) Fourier transform in the transverse plane, and $T(k_x,k_y,z) = \exp [i k z -(k_x^2+k_y^2)z/2k \;]$, with 
$k_x$ and $k_y$ wave numbers in the $x$ and $y$ directions, is the angular-spectrum transfer function \cite{goodman}.
The resulting intensity (upper) and phase (lower row) 
distributions for diffracted BG (second) and LG (fourth column) 
modes with azimuthal index $l_0=1$ are presented in Fig. \ref{InPhase}, in comparison to those of the incident BG (first) and LG (third column) modes.

Clearly, the obstacle significantly perturbs both types of input modes. However, both, the intensity and the phase distributions of the BG mode are less affected by the obstruction than those of the LG mode. This  enhanced stability of BG modes is known \cite{BouchalOC1998} and attributed to their ``healing'' property (which LG modes with $p=0$ lack), that is the feature to restore their spatial structure following disturbances. It is worth pointing out that the concept of self healing is not quantitative and purely based on a visual inspection of a finite region around the beam axis. The effect of the obstacle is still present in the transverse plane, but it has been pushed away from the observation window. This is easy to understand if we think of BG modes as superpositions of waves propagating on a cone. 

A modification of the intensity and phase distributions of diffracted 
modes as observed in Fig.~\ref{InPhase} can be regarded as the manifestation of diffraction-induced coupling of the spatial modes of the incident wave to other spatial modes. In other words, the diffracted mode \eqref{norm_psi} 
corresponds to a normalized single photon diffracted state
\begin{equation}
|\psi_{LG}\rangle=\sum_{pl}c_{pl}(z)|p,l\rangle,
\label{diff_LG}
\end{equation}  
for the LG mode, and to
\begin{equation}
|\psi_{BG}\rangle=\sum_{l}\int dk_\rho c_{l}(z,k_\rho)|k_\rho,l\rangle,
\label{diff_BG}
\end{equation}  
for a BG mode, with $|p,l\rangle$ and $|k_\rho,l\rangle$ single photon states of the modes $u^{pl}_{LG}(x,y,z)$ and $u^{k_\rho l}_{BG}(x,y,z)$, respectively. The latter, unperturbed mode functions are known exactly \cite{AllenPRA1992,GoriOptComm1987}, but they can also be easily assessed numerically, by plugging Eqs. \eqref{LG} and \eqref{BG} into the angular-spectrum propagator \eqref{angspec}.
From Eqs.~\eqref{diff_LG} and \eqref{diff_BG} it is clear that diffraction introduces crosstalk among different OAM modes \cite{Anguita:08,TylerOptLett09}.
The expansion coefficients are then given by the  inner products of the diffracted field and the unperturbed mode functions, 
\begin{subequations}
\begin{eqnarray}
&c_{pl}(z)&\!= \!\int\! \int\! dx dy u_{LG}^{pl*}(x,y,z) \psi_{LG}(x,y,z), 
\label{coeffa} \\
&c_{l}(z,k_\rho)&=\!\int \!\int\! dx dy u_{BG}^{k_\rho l*}(x,y,z) \psi_{BG}(x,y,z).
\label{coeffb}
\end{eqnarray}
\label{coeffs}
\end{subequations}
We point out that the inner products \eqref{coeffs} are invariant under translations along the $z$-axis. This can be proven \cite{ChuOptExpr2014} using the Plancherel theorem \cite{goodman} and the fact that the $z$-propagation is determined only by the angular-spectrum transfer function [see Eq. \eqref{angspec}]. Therefore, henceforth we will omit the $z$-dependence of the expansion coefficients.

A standard way to analyze crosstalk is through the assessment of the expansion coefficients in Eq. \eqref{coeffs}. However, even without addressing the individual coupling amplitudes $c_{pl}$, $c_{l}(k_\rho)$, we can build some qualitative understanding of the presence of different OAM modes in the diffracted wave.  
In the next section we will see how the scattering of the input mode into a superposition of many OAM modes induces a non zero mutual overlap of the diffracted waves, and thereby affects the output state entanglement.

Let us therefore examine the bottom row of Fig. \ref{InPhase}  where the phase profiles of the incident (azimuthal index $l_0=1$) and diffracted (in general, arbitrary $l$-values) waves are depicted.
We identify the $l_0 = 1$ character of the incident modes in Figs. \ref{InPhase}(e) and (g) by their only phase singularity at the origin, associated with a topological charge one. Inspection of the phase profile of the diffracted beam shows that, notwithstanding some phase distortions, the diffracted BG mode [Fig. \ref{InPhase}(f)] still exhibits the very sole phase singularity at the origin.
This suggests that the azimuthal index is almost preserved, or that scattering to other OAM modes is weak. In contrast, the appearance of multiple phase singularities for the diffracted LG mode -- for example, along the positive $x$ axis -- is evident [Fig.~\ref{InPhase}(h)]. This suggests that this phase structure belongs to a superposition of multiple OAM modes.

\section{Diffraction of a biphoton on two opaque screens}
\label{EnCo}

We can now import our above results for the diffraction of a single input mode, to quantify the entanglement decay of an OAM bi-photon input state when each photon 
is diffracted upon an obstacle.

While this scenario is inspired by the experiment in \cite{McLarenNatPhys2014}, the physics here considered is different inasmuch as the experiment quantified the output state entanglement as a function of the distance between obstacle (centered at the optical axis) and detector. For distances smaller than the self-healing length of the encoding BG modes, measurement noise obscures the output state entanglement in \citep{McLarenNatPhys2014}. In contrast, in our present set-up the distance between obstacle and detector is invariant, and we instead quantify entanglement reduction induced by transverse displacements of the obstacle, which leads to non trivial scattering into different OAM modes upon transmission. Therefore, the here observed entanglement reduction is due to an actual perturbation of the transmitted state rather than to a smooth modulation of the signal to noise ratio.
\subsection{Setup}
\label{Setup}

\begin{figure}
\includegraphics[width=1\linewidth]{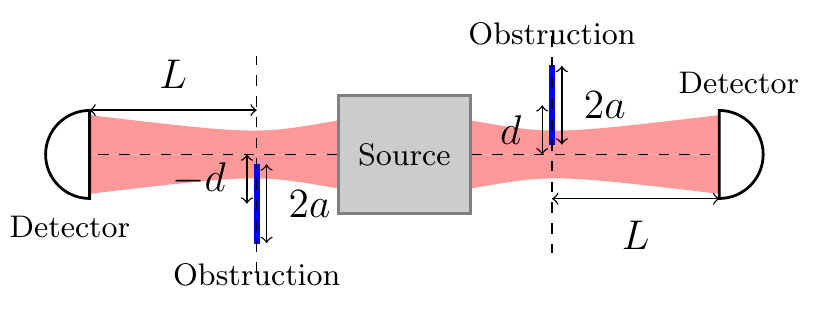}
\caption{Sketch of the setup. A source produces a pair of maximally-entangled twisted photons. The two photons propagate in opposite directions; at their beam waists ($z=0$) they are diffracted by circular obstructions of radius $a$. The two obstacles are displaced by $\pm d$ with respect to the beam axis. Finally, the photons are detected at a distance $L$ from the obstacles.}
\label{entfig}
\end{figure}

More specifically, we consider the setting in Fig. \ref{entfig}: 
A source generates  pairs of single photon 
excitations of LG or BG modes, which are Bell state (i.e., maximally) entangled in their OAM,
\begin{equation}
\label{psi0}
\ket{\Psi_0} = \frac{1}{\sqrt{2}}(\ket{l_0,-l_0} + \ket{-l_0,l_0}),
\end{equation}
with $\ket{\pm l_0}$ the shorthand notation for either $\ket{0,\pm l_0}$ (LG mode) or $\ket{\kappa,\pm l_0}$ (BG mode).
To minimize the notational overhead, we assume that each of the 
photons is diffracted by a circular screen, and that both screens have identical radii and are placed with opposite offsets with respect to the optical axis. 
For simplicity and without loss of the essential physics, in our calculations we set both screens at $z=0$.\footnote{Finite distances $z$ would lead to the appearance of $z$-dependent propagation phases (see Sec.~\ref{sec:basic}) in the obstacle plane, resulting, e.g., in rotation of the phase distributions in Fig.~\ref{InPhase}(e,g). However, this rotation does not affect entanglement.}
Finally, the biphoton state is calculated at a distance $L$ from the obstacle planes.

Under the transformation \eqref{bc} of each of the modes acting as carriers of the single photon components $|\pm l_0\rangle$ of the twin-photon state \eqref{psi0}, the 
input state's underlying mode structure is 
mapped on the (normalized, according to Eq. \eqref{norm_psi}) diffracted output mode
\begin{eqnarray}
\Psi(\mathbf{r}_1,\mathbf{r}_2;L)&= \frac{1}{\sqrt{2}}  \left [\psi_{l_0}^d(\mathbf{r}_1,L)\psi_{-l_0}^{-d}(\mathbf{r}_2,L) \right . \nonumber\\
& \left . +\psi_{-l_0}^d(\mathbf{r}_1,L)\psi_{l_0}^{-d}(\mathbf{r}_2,L) \right ],
\label{diff_2photon}
\end{eqnarray}
where $\psi_{l}^{\pm d}(\mathbf{r}_i, L)$ [$\mathbf{r}_i=(x_i,y_i)$ and $i=1,2$] represents the constituent single modes' diffractive image (at the distance $L$ from the obstruction). Accordingly, by generalization of Eqs. (\ref{diff_LG},\ref{diff_BG}), the diffracted bi-photon output state reads
\begin{equation}
|\Psi(L)\rangle\!=\!\begin{cases}
\sum_{p_1,l_1,p_2,l_2}c_{p_1l_1,p_2l_2}|p_1,l_1;p_2,l_2\rangle,\\\\
\sum_{l_1,l_2}\int dk^\prime_{\rho}dk^{\prime\prime}_{\rho}c_{l_1l_2}(k^\prime_{\rho},k^{\prime\prime}_{\rho})|k^\prime_{\rho},l_1;k^{\prime\prime}_{\rho},l_2\rangle,
\end{cases}
\label{Psi_L}
\end{equation}
for OAM encoding in the LG (first line) or in the BG (second line) basis, respectively.
where the first and second lines correspond to the expansion in the LG and BG bases, respectively, and the dependence on $L$ in the right hand side of Eq. (\ref{Psi_L}) is incorporated solely in the basis states. The latter dependence, however, cannot affect entanglement, which is encoded in the expansion coefficients. Thus, entanglement remains invariant under $z$-translations. Therefore, we can lighten the notation and drop the propagation distance $L$ in all subsequent expressions.

\subsection{Entanglement of the output state}
\label{sec:ent}
\begin{figure}[t!]
\includegraphics[width=0.9\linewidth]{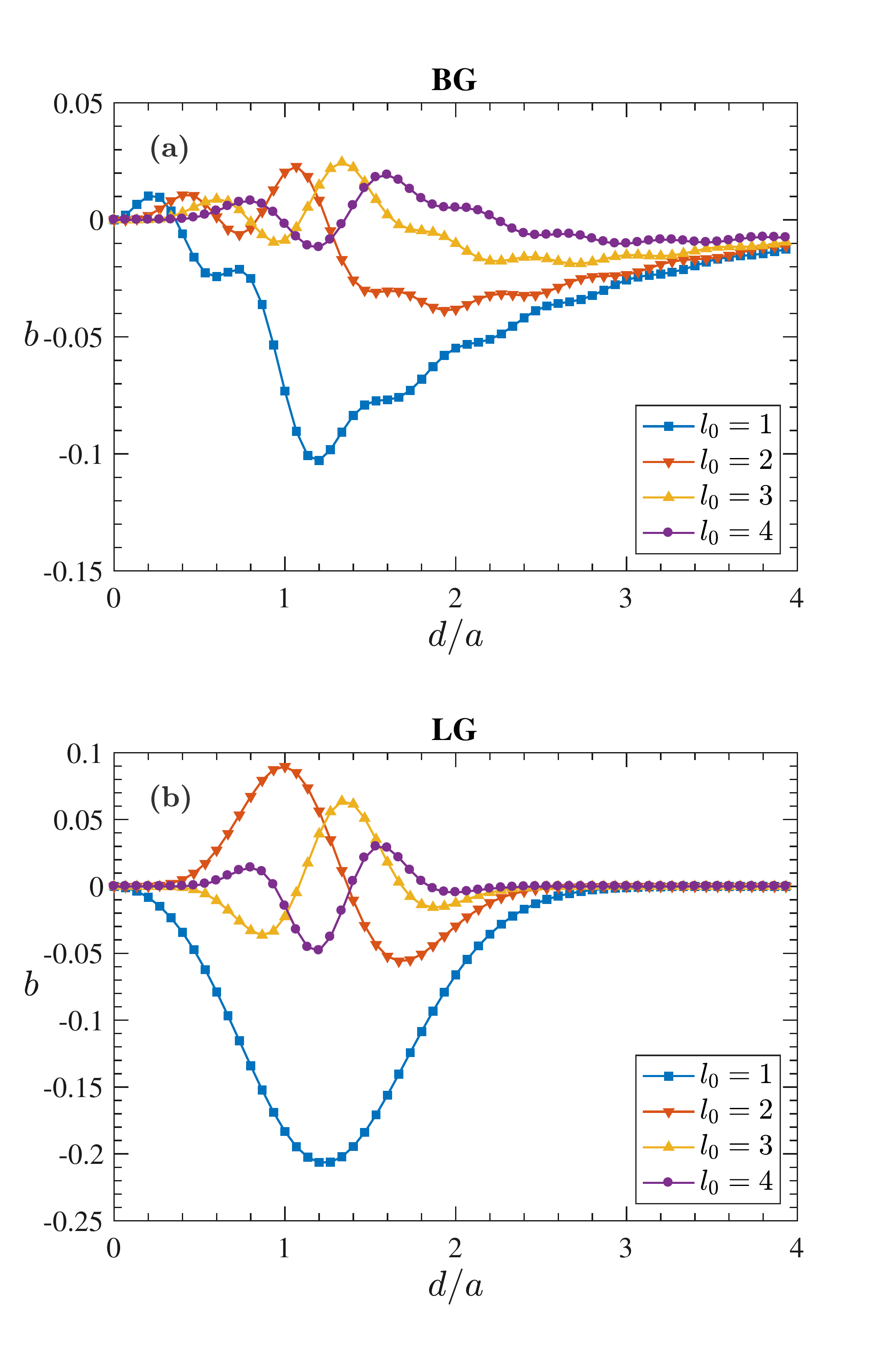}
\caption{(Color online) Mutual overlap $b$ versus the relative displacement $d/a$ (where $a=200\;\mu$m) for (a) BG modes with $l_0=1,2,3,4$ and $w_{BG} = 1$ mm, and (b) for  LG modes with the same azimuthal numbers and $w_{LG}=226, 160, 130, 113\;\mu$m, respectively, where variable beam widths ensure screening off $l_0$-independent structural elements of the LG beams [see the discussion following Eq. \eqref{BG}].}
\label{bplot}
\end{figure}
\begin{figure}[t!]
\includegraphics[width=0.9\linewidth]{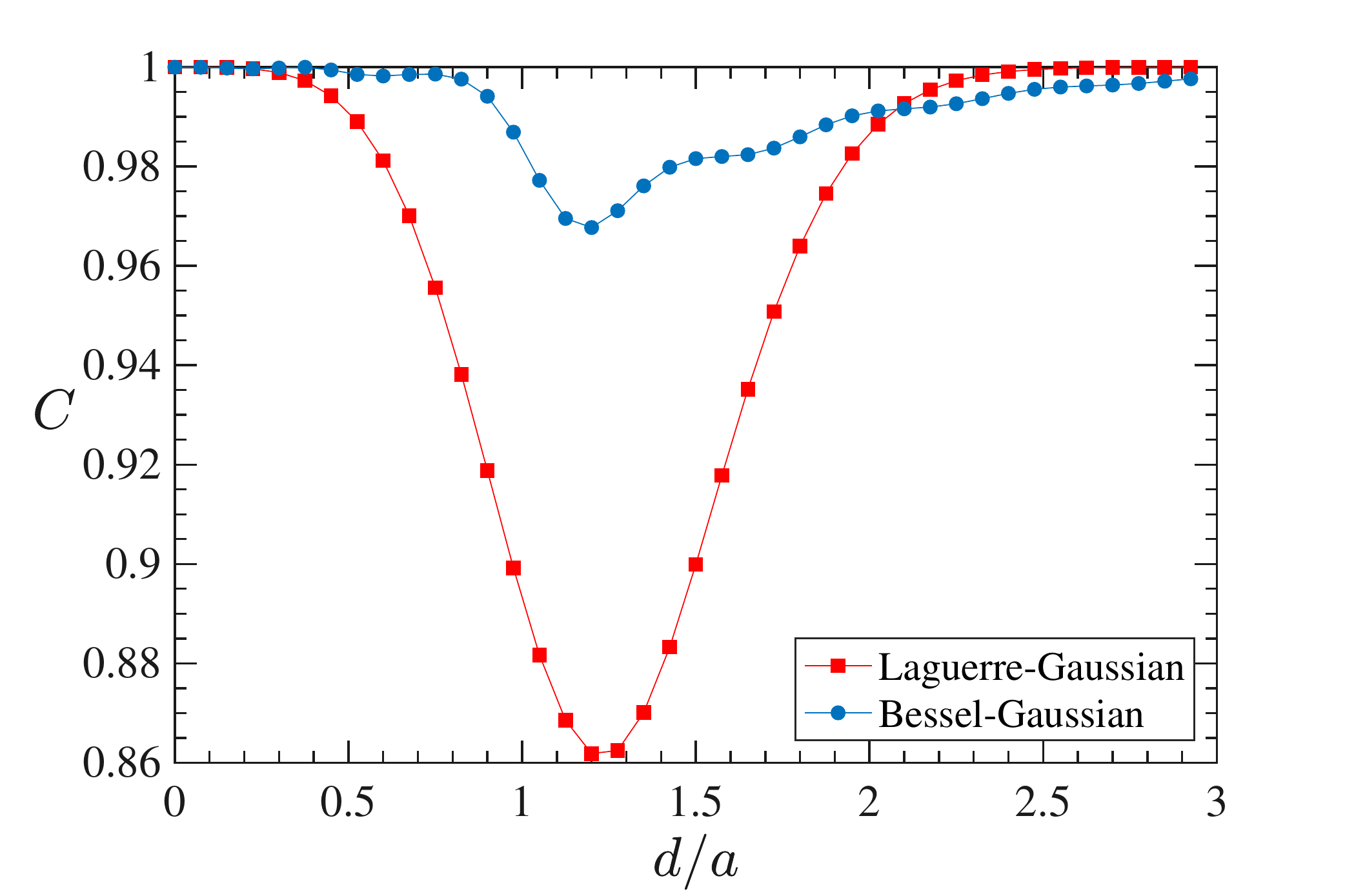}
\caption{(Color online) Concurrence $C(|\Psi\rangle)$ versus the relative displacement $d/a$ ($a=200\;\mu$m), for the diffracted twin-photon state \rref{psi0} of LG (red squares) and BG (blue dots) modes with $l_0 = 1$. Beam waists were set to $w_{LG} = 226\;\mu$m and $w_{BG} = 1$ mm .}
\label{Cbglg}
\end{figure}

Given the above, we can now proceed to quantify the diffracted bi-photon output state's entanglement, directly from its transverse position representation \eqref{diff_2photon}. For this purpose we employ the higher dimensional generalization \cite{RungtaPRA2001, GuoQuantInfProc2013} of concurrence \cite{WoottersPRL1998}
\begin{equation}
\label{C}
C(|\Psi\rangle) = \sqrt{2\left(1 - \tr[\varrho^2]\right)},
\end{equation}
where $\varrho=\tr_2\left[|\Psi\rangle\langle \Psi|\right]$ is the reduced density matrix of either one of the entangled 
photons, after tracing out the other. This trace is performed on the output state's two photons density matrix
\begin{equation} 
\sigma(\mathbf{r}_1,\mathbf{r}_2;\mathbf{r}^{\prime}_1,\mathbf{r}^{\prime}_2)=\Psi^*(\mathbf{r}_1,\mathbf{r}_2)\Psi(\mathbf{r}^{\prime}_1,\mathbf{r}^{\prime}_2) \, .
\label{sigma}
\end{equation}
by integration over the transverse coordinates of the second photon, to obtain 
\begin{align}
\varrho(\mathbf{r}_1,\mathbf{r}_1^\prime) & = \int d^2r_2 \sigma(\mathbf{r}_1,\mathbf{r}_2;\mathbf{r}_1^\prime,\mathbf{r}_2)\nonumber\\
&=\frac{1}{2}\left [ \psi_{l_0}^d(\mathbf{r}_1)\psi^{d\,*}_{l_0}(\mathbf{r}_1^\prime)+b \psi_{l_0}^d(\mathbf{r}_1)\psi^{d\,*}_{-l_0}(\mathbf{r}_1^\prime) \right. \label{reduced_dm}
\\ 
&\left.+ b \psi^d_{-l_0}(\mathbf{r}_1)\psi^{d\,*}_{l_0}(\mathbf{r}_1^\prime)+ \psi^d_{-l_0}(\mathbf{r}_1)\psi^{d\,*}_{-l_0}(\mathbf{r}_1^\prime)\right ]. \nonumber
\end{align}
In~\eqref{reduced_dm} we introduced the (real) parameter,
\begin{equation}
b \equiv \int d^2 r \psi^{d\,*}_{-l_0}(\mathbf{r})\psi^d_{l_0}(\mathbf{r})\, ,
\label{crosstalk2}
\end{equation}
which is the mutual overlap between the diffracted fields $\psi^d_{- l_0}(\mathbf{r})$ and $\psi^d_{+ l_0}(\mathbf{r})$. Scattering of the input fields $u_{\pm l_0}(\mathbf{r},z=0)$ into superpositions of OAM modes (see Fig.~\ref{InPhase}) results in a nonzero value of $b$ if and only if some of the modes in the diffracted fields are common. By virtue of Eq. (\ref{crosstalk2}), 
$0\leq |b| \leq 1$, with the upper bound attained in the degenerate 
limit $-l_0=l_0=0$, where, however, 
$|\Psi_0\rangle$ 
reduces to a product state. 

\begin{figure*}[t!]
\includegraphics[width=0.9\linewidth]{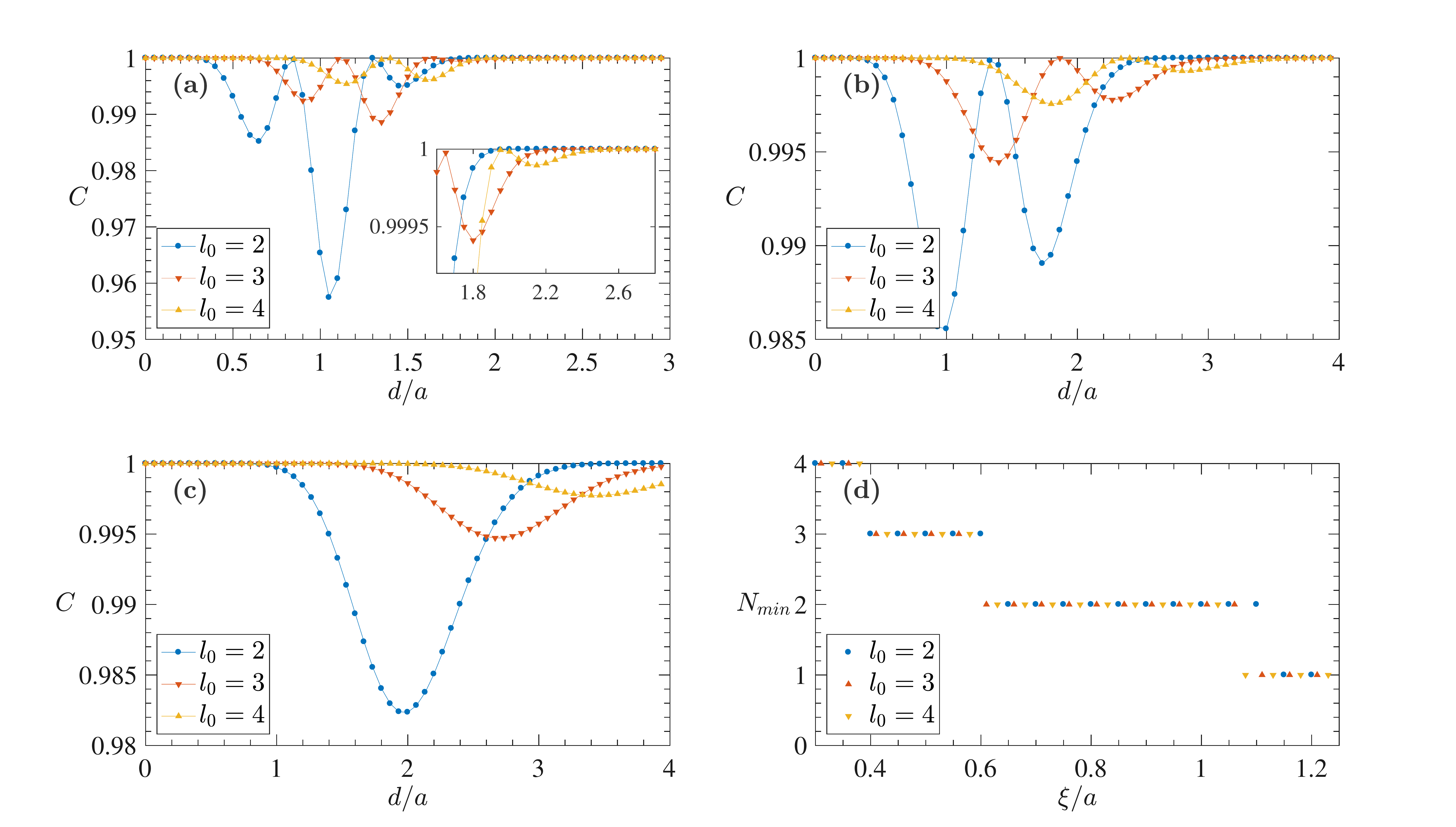}
\caption{(Color online) (a-c) Concurrence $C(|\Psi\rangle)$ versus the relative displacement $d/a$ ($a = 200\;\mu$m), for the diffracted twin-photon state \eqref{psi0} of LG modes with $l_0 = 2,3,4$. Panels (a,b,c) correspond to three distinct values of the ratio $\xi/a = 0.4, 0.75, 1.2$, respectively. (d) Number of minima of the concurrence $N_{min}$ versus $\xi(l_0)/a$ [see Eq. \eqref{xi}], for  $l_0 = 2,3,4$. }
\label{Cxi}
\end{figure*}

Several examples of the mutual overlap $b$ as a function of the relative displacement $d/a$ are shown in Fig. \ref{bplot}. We first note that $b=0$ if $d=0$. In this case, the mutual overlap vanishes, since the perturbation caused by the obstacle leaves the cylindrical symmetry of diffracted waves intact.
Furthermore, $b\to 0$ when the screens' displacement is large compared to the essential support of the beam, due to the trivial reason that the  screens' impact turns negligible in this limit.
For intermediate displacements of the obstructions, interference of the diffracted modes results in oscillations of $b$, with amplitudes which are smaller for BG than for LG 
encoding, for a given $l_0$, and progressively decrease with increasing OAM, for both sets of modes. 
One can see a progression in the oscillation of the mutual overlap in the LG case. The number of peaks and dips is equal to the OAM index $l_0$ (see also Sec. \ref{sec:phase_corr}). The oscillations are modulated by an envelope function, whose amplitude and width gradually decrease with increasing $l_0$. In the case of $l_0=1$, the dip in the oscillation more or less coincides with the peak of the envelope function. Therefore its amplitude seems disproportionately large compared to those for higher orders. In the BG case, the shapes of the curves are more complex. There are still oscillations, but their modulation is more irregular. Nevertheless, a similar progression causes the amplitudes of these curves to decrease gradually for increasing $l_0$.

The significance of the mutual overlap $b$ is that it 
uniquely determines the diffraction-induced entanglement decay.
Indeed, using the normalization of $\psi^d_{\pm l_0}({\bf r})$,
 as well as Eqs. (\ref{reduced_dm}) and (\ref{crosstalk2}), we find
\begin{align}
\label{redpur}
\tr(\varrho^2) &= \int d^2 r \int d^2 r^\prime \varrho(\mathbf{r},\mathbf{r}^\prime) \varrho(\mathbf{r}^\prime,\mathbf{r}) \\
&=\frac{1}{2}(1 + 6b^2 + b^4), \nonumber
\end{align}
which leads to an explicit expression for the output state's concurrence \eqref{C}, as the main result of this work:
\begin{equation}
C(|\Psi\rangle)=\sqrt{1-6b^2-b^4}.
\label{Can}
\end{equation}
However, $b$ is not an independent parameter but a function of the beam waist, of the size and displacement of the screen, and of the azimuthal index.

\subsection{Entanglement loss upon diffraction}
\label{sec:ent-res}

Let us now assess the entanglement decay featured by the diffracted image of \rref{psi0} -- for encoding in $l_0 = 1$. This is the case for which we have the maximum mutual overlap 
(see Fig. \ref{bplot}).
Fig. \ref{Cbglg} shows the dependence of $C(|\Psi\rangle)$ on the relative displacement $d/a$. Comparison to Fig. \ref{bplot} shows that the behaviour of $C(|\Psi\rangle)$ follows directly from that of $b$ (see Fig. \ref{bplot}), and that, in particular, OAM entanglement is more robust in the BG as compared to the LG basis, for most values of $d/a$. Only for large displacements does LG offer a tiny advantage, since the multi-ring intensity profile of BG modes leads to some residual overlap even at large $d/a$. This multi-ring structure is also at the origin of the 
modulation of the BG mode's output concurrence with the displacement, and of the underlying mutual overlap behaviour depicted in
Fig. \ref{bplot}(a).
Minimal output concurrence is observed, for both encodings, 
at $d/a\approx 1.2$, corresponds to stationary points of the mutual overlap $b$ in Fig.~\ref{bplot}, and is due to the obstacles' overlaps with the maxima of the intensity distribution of the input beams (e.g., the LG input mode with $l_0=1$ and $p=0$ has an intensity maximum 
concentrated along a single ring with radius $\approx w_{LG}$, and $C(|\Psi\rangle)$ exhibits a minimum precisely when the obstacle is placed at the 
corresponding position).

\subsection{Entanglement and the phase correlation length}
\label{sec:phase_corr}

The situation is slightly more complicated for LG-encoded OAM entangled input states \eqref{psi0} with $l_0\geq 2$. While an LG mode's intensity profile exhibits only one bright ring also for $l_0\geq 2$, the mutual overlap $b$ oscillates, as shown in Fig. \ref{bplot}(b), and so does the output concurrence in Fig. \ref{Cxi}, due to \eqref{Can}. We therefore need to resort to the output state's phase distribution, and, in particular, to its {\it phase 
correlation length} $\xi(l_0)$ \cite{LeonhardPRA2015}, which accounts for both, its phase and intensity profiles.

For LG modes with $p=0$ and $|l_0|\geq 2$, $\xi(l_0)$ reads \cite{LeonhardPRA2015}
\begin{equation}
\xi(l_0) = \frac{w_{LG}}{\sqrt{2}}\sin\left(\frac{\pi}{2|l_0|}\right)\frac{\Gamma(3/2 +|l_0|)}{\Gamma(1+|l_0|)}\, ,
\label{xi}
\end{equation}
and it was shown that the entanglement decay experienced by LG-encoded OAM entangled states in a weakly turbulent atmosphere 
is a universal function of the ratio $\xi(l_0)/r_0$, where $r_0$ is the turbulence coherence length (the typical length scale on which turbulence-induced phase errors are correlated \cite{FRIED:65}).
In our present scenario, a natural analog of $r_0$ is the radius $a$ of the obstacle, and the curves in Fig.~\ref{Cxi} are obtained for 
distinct values of the ratio $\xi(l_0)/a$ for each of the panels (a-c).
While there is no apparent regularity of the behaviour 
of $C(|\Psi\rangle)$ in either one of the panels (a-c), the results can be classified by one specific feature -- the number of minima $N_{min}$ of $C(|\Psi\rangle)$. 
To demonstrate this, panel (d) correlates $N_{min}$ with the ratio $\xi(l_0)/a$: Indeed, for small values $\xi(l_0)/a\lesssim 0.4$, $N_{min}=4$, and with increasing $\xi/a$, $N_{min}$ steadily decreases by one, at $\xi(l_0)/a\simeq 0.6$ and $\xi(l_0)/a\simeq 1.07$.

\section{Conclusion}
\label{conc}
    
In this work, we studied the effect of the diffraction of a singlet type biphoton OAM state, initially encoded either in LG or in BG modes, on the entanglement content of such states. Using methods of Fourier optics we inferred the diffracted twin-photon state, taking into account the size and shift of the obstacle with respect to the beam axis, as well as different initial azimuthal quantum numbers $l_0$ of the modes.

We derived an analytical formula for the concurrence of the diffracted state, which is an infinite dimensional, pure state. This formula depends only on one parameter -- the scattering-induced mutual overlap between the two diffracted waves which stem from the OAM modes with indices $l_0$ and $-l_0$, respectively. This result, in particular, shows that peculiarities of spatial distributions of BG and LG modes lead to different mutual overlaps of diffracted waves, which, eventually, results in stronger robustness of entanglement for BG modes. 
Thereby, our theoretical findings complement the experimental observations of \cite{McLarenNatPhys2014}.

This work corroborates that scattering into superpositions of OAM modes underlies the behavior of entanglement under deterministic perturbations, such as diffraction, as much as it does under random disturbances, such as weak atmospheric turbulence \cite{LeonhardPRA2015}. Therefore, our results suggest 
a stronger resilience in atmospheric turbulence of photonic OAM entanglement encoded into BG than into LG modes -- a statement 
which will be interesting to verify in the future.   
Another possible research direction will be to interpret the diffraction-induced spreading of the OAM spectrum in terms of the uncertainty principle for angular position and angular momentum \cite{franke_arnold_2004} and to establish a connection between angular uncertainty and OAM entanglement loss.

\acknowledgments
We would like to thank A. Forbes for illuminating and enjoyable discussions. G.S., V.N.S. and A.B. acknowledge support by Deutsche Forschungsgemeinschaft under grant DFG BU 1337/17-1.

\bibliography{PaperBiblio}
\end{document}